\documentclass{article}
\usepackage{spconf,amsmath,graphicx}
\usepackage{amssymb}
\usepackage{amsmath}
\usepackage[colorinlistoftodos, disable]{todonotes}
\usepackage{booktabs} 
\usepackage{hyperref}
\usepackage{bm}
\usepackage{comment}

\title{DDSP-based Neural Waveform Synthesis of Polyphonic Guitar Performance from String-wise MIDI Input}
\name{Nicolas Jonason$^1$, Xin Wang$^2$, Erica Cooper$^2$, Lauri Juvela$^3$, Bob L. T. Sturm$^1$, Junichi Yamagishi$^2$}
\address{$^1$KTH Royal Institute of Technology, Sweden
$^2$National Institute of Informatics, Japan \\
$^3$Aalto University, Finland}

\begin{document}
\ninept
\maketitle
\begin{abstract}
We explore the use of neural synthesis for acoustic guitar from string-wise MIDI input. We propose four different systems and compare them with both objective metrics and subjective evaluation against natural audio and a sample-based baseline.
We iteratively develop these four systems by making various considerations on the architecture and intermediate tasks, such as predicting pitch and loudness control features.
We find that formulating the control feature prediction task as a classification task rather than a regression task yields better results. Furthermore, we find that our simplest proposed system, which directly predicts synthesis parameters from MIDI input performs the best out of the four proposed systems. Audio examples are available at \href{https://erl-j.github.io/neural-guitar-web-supplement/}{https://erl-j.github.io/neural-guitar-web-supplement}.
\end{abstract}

\section{Introduction}

The synthesis of expressive and realistic guitar performance has been attempted in several ways. One is sample-based synthesis \cite{amplesound_ample_2015}, which requires a database of purpose-made sample recordings. Another way is physical modeling synthesis \cite{laurson_methods_2001}, which requires solving systems of partial differential equations modeling the entire guitar. This paper considers a third approach: neural synthesis, a data-driven approach that does not require a purpose-made sample library or the specification of the physics of the entire instrument.

One family of neural synthesis techniques, termed Differentiable Digital Signal Processing (DDSP) \cite{engel_ddsp_2019}, involves integrating digital signal processors into neural networks. One particular DDSP configuration combines harmonic oscillators, noise filtering and a trainable reverb \cite{engel_ddsp_2019}. 
This has been used to model instruments such as violin (monophonic) \cite{engel_ddsp_2019}, various wind instruments \cite{jonason_neural_2022,
wu_midi-ddsp_2021} and piano \cite{renault_ddsp-piano_2023} with high quality from small amounts of training data.
While Engel \MakeLowercase{\textit{et al.}} \cite{engel_ddsp_2019} originally controlled the networks with pitch and loudness, later work extended these models to other forms of input such as higher level expression features \cite{wu_midi-ddsp_2021} and MIDI input \cite{renault_ddsp-piano_2023,wu_midi-ddsp_2021,castellon_towards_2020,jonason_control-synthesis_2020}.

Other work has shown that one can train polyphonic DDSP models, where \textit{voices} are rendered in parallel and then summed, by optimizing a spectral loss between the ground truth polyphonic audio and the mixture of synthesized voices  \cite{renault_ddsp-piano_2023,kawamura_differentiable_2022}.

Guitar poses several challenges from a neural synthesis perspective. First, it is a polyphonic instrument, which means that multiple voices can be active at the same time each with varying pitch. Guitar performance involves idiosyncrasies such as bending strings, sliding, and other articulations. In the case of bends, pitch is not discrete; and in the case of playing legato, pitch can vary within a single string excitation. Furthermore, the type, angle and location of the excitation of a string each affects the resulting sound.
Unlike piano performance, where detailed transcriptions can be digitally captured (e.g., by a Disklavier), obtaining objective and detailed ground truth transcription for guitar performances is difficult. For example, strummed chords often include muted strings that are sometimes omitted from the transcription \cite{xi_guitarset_2018}.

Our work adapts and develops the {\em control-synthesis approach}
\cite{jonason_control-synthesis_2020} to the synthesis of acoustic guitar performance. While some work has explored neural synthesis of single guitar notes \cite{engel_neural_2017, community_words_2022, engel_gansynth_2018}, our work is the first to use neural networks to generate performances of guitar from string-wise MIDI input.
We train four different neural synthesizers on hexaphonic and microphone recordings of acoustic guitar performances from string-wise MIDI input, and compare them with both objective metrics and
subjective evaluation against natural audio and a sample-based base-
line. We start with a basic model that treats control feature prediction as a regression task. We find better performance results from treating control feature prediction as a classification task instead. We then experiment with joint training of control and synthesis sub-models, which further improves performance. Finally, we experiment with a unified architecture that merges control synthesis sub-modules and simplifies training.
Audio examples are available at \href{https://erl-j.github.io/neural-guitar-web-supplement/}{https://erl-j.github.io/neural-guitar-web-supplement}.

\vspace{-1mm}\section{Proposed systems}

\begin{figure}
    \centering
\includegraphics[width=\columnwidth]{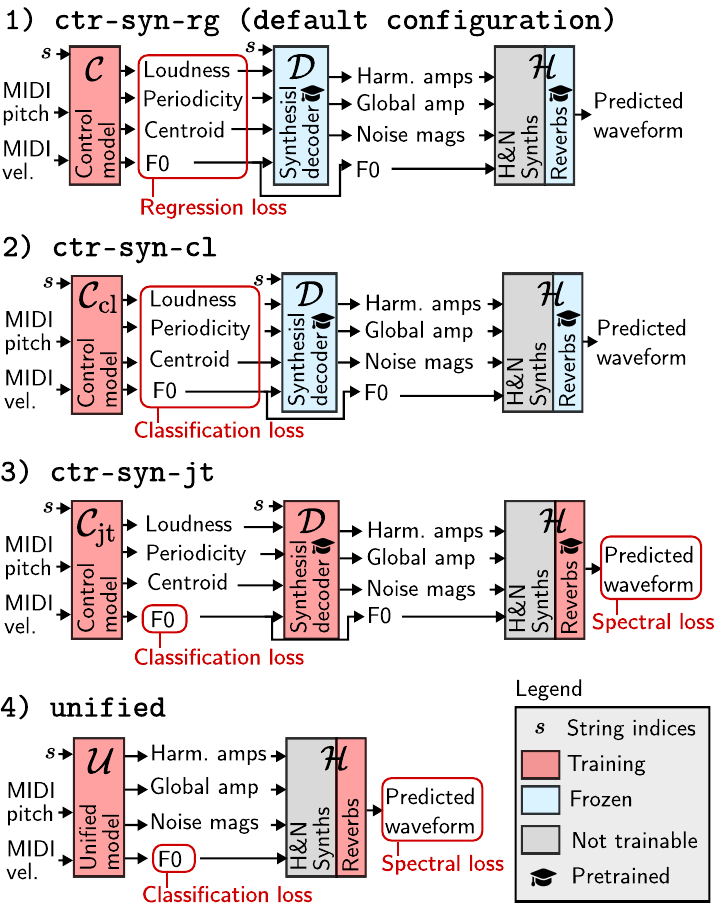}
    \vspace{-5mm}
    \caption{Overview of the four proposed systems. \texttt{ctr-syn-rg} is a control-synthesis model with control feature regression. \texttt{ctr-syn-cl} is a control-synthesis model with control feature classification. \texttt{ctr-syn-jt} uses joint training of control and synthesis submodules. \texttt{unified} merges the control model and synthesis decoder into a single network.}
    \label{fig:systems}
    \vspace{-5mm}
\end{figure}

We now introduce our proposed systems, starting by presenting the default configuration and then its subsequent modifications. Figure \ref{fig:systems} shows the four configurations.

\subsection{Default configuration}

Our default configuration (\texttt{ctr-syn-rg}) uses a control-synthesis architecture \cite{jonason_control-synthesis_2020}.
The control model $\mathcal{C}$, written as
\(
\bm{\hat{f_0}},\bm{\hat{l}},\bm{\hat{p}},\bm{\hat{c}} = \mathcal{C}(X_{\text{pitch}}, X_{\text{vel}}, s)
\), 
takes string-wise MIDI input consisting of one-hot quantized pitch \(X_{\text{pitch}} \in \{0,1\}^{(n_{\text{strings}}, n_{\text{frames}}, n_{\text{pitch bins}})}\) and one-hot quantized velocity \(X_{\text{vel}} \in \{0,1\}^{(n_{\text{strings}}, n_{\text{frames}}, n_{\text{vel bins}})}\) as well as a vector of one-hot encoded string indices \(s \in \{0,1\}^{(n_{\text{strings}})}\).
Its output consists of predictions of four control features for each string: the fundamental frequency (F0) \(\bm{f_0}\), the loudness \(\bm{l}\), the periodicity \(\bm{p}\), and the spectral centroid \(\bm{c}\) where \(\bm{f_0},\bm{l},\bm{p},\bm{c} \in \mathbb{R}^{(n_{\text{strings}}, n_{\text{frames}})}\).
We define $n_{\text{pitch bins}}=305$, $n_{\text{vel bins}}=64$,
and $n_{\text{strings}}=6$, whereas $n_{\text{frames}}$ depends
on the feature frame rate and render duration.
We include periodicity in order to help the synthesis model distinguish between tonal and non-tonal sections \cite{jonason_neural_2022}.
We include spectral centroid to provide timbral information to the synthesis model.

The synthesis model consists of two parts, a synthesis decoder and a harmonic+noise+reverb synthesizer. The synthesis decoder \(\mathcal{D}\), written as
\(
    H, \bm{a}, N = \mathcal{D}(\bm{\hat{f_0}},\bm{\hat{l}},\bm{\hat{p}},\bm{\hat{c}}, s)
\),
takes string-wise predicted control features from $\mathcal{C}$ and generates three synthesis parameters, consisting of harmonic amplitudes \(H \in \mathbb{R}^{(n_{\text{strings}}, n_{\text{frames}}, n_{\text{harmonics}})}\), global harmonic amplitude \(\bm{a} \in \mathbb{R}^{(n_{\text{strings}}, n_{\text{frames}})}\) and filtered noise band amplitudes \(N \in \mathbb{R}^{(n_{\text{strings}}, n_{\text{frames}}, n_{\text{noise bands}})}\). 
These synthesis parameters are then fed to a 6-voice harmonic + noise + reverb synthesizer $\mathcal{H}$,
the outputs of which are summed to 
predict the final waveform \(y \in \mathbb{R}^{(n_{\text{samples}})}\).

\begin{figure*}[t]
    \centering
    \includegraphics[width=0.80\textwidth]{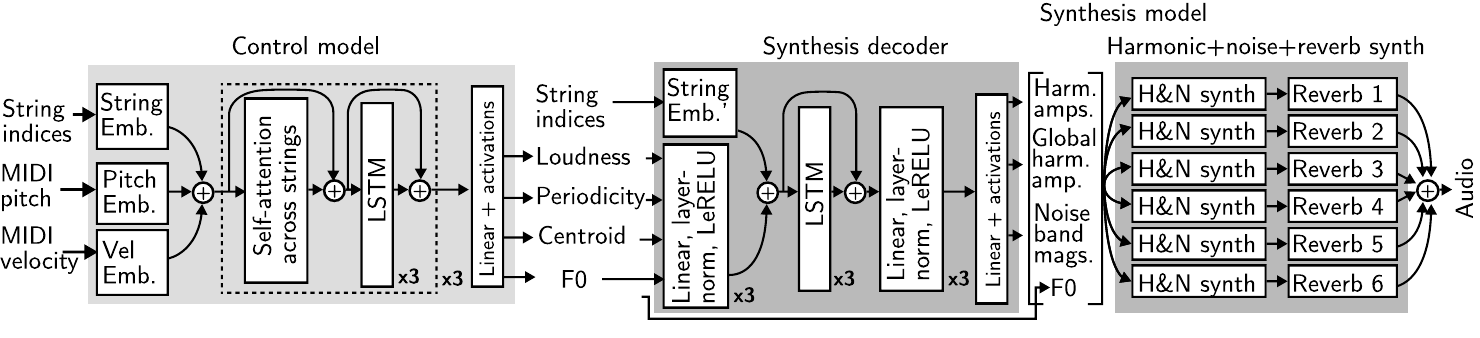}
    \caption{Details of the control-synthesis architecture}
    \label{fig:default}
    \vspace{-5mm}
\end{figure*}

Figure \ref{fig:default} details the control-synthesis architecture.
The control model $\mathcal{C}$ consists of a neural network combining bi-directional long short-term memory (LSTM) \cite{hochreiter1997long} operating across the time dimension and self-attention \cite{vaswani_attention_2017} operating across the string dimension to account for inter-string dependencies.
The synthesis decoder $\mathcal{D}$ uses a bi-directional LSTM-based neural network \cite{engel_ddsp_2019}.
The harmonic + noise synthesizer uses 128 harmonics and 128 noise filter bands, with the same design as in \cite{engel_ddsp_2019}.
While acoustic guitar does exhibit inharmonicity due to string stiffness, the inharmonicity is difficult for the average listener to perceive \cite{karjalainen_is_2005}, and we leave the inclusion of inharmonicity to future work.
We use one trainable reverb per string to account for differences in the position of each string in relation to the guitar body and the microphone. Each reverb has a $0.25$-second trainable impulse response.
With the exception of the self-attention across strings, all network layers process the strings independently. 
The control model and synthesis decoder use a hidden-layer size of $512$.
The control model, synthesis decoder and trainable reverb have 53.7M, 18.2M, and 72k parameters respectively, totalling 72M parameters.

In the default configuration, we train the control and synthesis models separately. First, we train the synthesis model to minimize $\mathcal{L}_{syn}=\text{MSSL}(y,y')$, i.e the multi-scale spectral loss (MSSL) \cite{wang_neural_2019, engel_ddsp_2019} between natural microphone audio $y$ and synthesized audio from ground truth control features $y'$. 
Second, we train the control model to predict control features from MIDI input. In line with earlier work \cite{jonason_neural_2022,wu_midi-ddsp_2021,castellon_towards_2020,jonason_control-synthesis_2020}, the default configuration treats the prediction of control features as a regression task, using a loss based on the mean squared error (MSE) of the predicted control features: 
$\mathcal{L}_\text{rg} =
   \mathcal{L}_{\bm{f_0}} +
   \mathcal{L}_{\bm{l}} +
   \mathcal{L}_{\bm{p}} +
   \mathcal{L}_{\bm{c}}$  where
\begin{align}
   \mathcal{L}_{\bm{f_0}} &= \sum_{i=1}^{n_{\text{strings}}}\sum_{t=1}^{T}
   (\bm{f_0}(i,t) - \bm{\hat{f_0}}(i,t))^2 \cdot \bm{l}(i,t) \cdot \bm{p}(i,t) \\
   \mathcal{L}_{\bm{l}} &= \sum_{i=1}^{n_{\text{strings}}}\sum_{t=1}^{T}
   (\bm{l}(i,t) - \bm{\hat{l}}(i,t))^2\\
   \mathcal{L}_{\bm{p}} &= \sum_{i=1}^{n_{\text{strings}}}\sum_{t=1}^{T}
   (\bm{p}(i,t) - \bm{\hat{p}}(i,t))^2 \cdot \bm{l}(i,t) \\
   \mathcal{L}_{\bm{c}} &=
   \sum_{i=1}^{n_{\text{strings}}}\sum_{t=1}^{T}
   (\bm{c}(i,t) - \bm{\hat{c}}(i,t))^2 \cdot \bm{l}(i,t)
\end{align}
where for string \(i \in [1,n_{\text{strings}}]\) and 
feature frame \(t\in [1,T]\) the target loudness $\bm{l}$ is used to weight the MSE
for $\bm{f_0}$ ,$\bm{p}$, $\bm{c}$ according to signal strength
thus discounting prediction errors in quieter sections.

Additionally, we discount the F0 loss in non-pitched sections, by weighting the $\bm{f_0}$ MSE by target periodicity.

\vspace{-2mm}
\subsection{Control feature prediction as a classification task}
After initial experiments with the default configuration produced poor results, we experimented with treating control feature prediction as a classification task rather than a regression task since this has been shown to yield better performance in other contexts \cite{zhang_improving_2022}.
We refer to this system as \texttt{ctr-syn-cl}.

We quantize and one-hot encode the control features with $n_\text{$f_0$ bins}=305$, and $K=64$ bins for $\bm{l}$, $\bm{p}$, $\bm{c}$. We denote the one-hot encoded features as \(F_0 \in \{0,1\}^{(n_{\text{strings}}, n_{\text{frames}}, n_{\text{$f_0$ bins}})}\), \(L, P, C \in \{0,1\}^{(n_{\text{strings}}, n_{\text{frames}}, K)}\).
The new control model, written as 
$\hat{F_0}, \hat{P}, \hat{C}, \hat{L} = \mathcal{C}_{\text{cl}}(X_{\text{pitch}}, X_{\text{vel}}, s)$, outputs probabilities for each control feature over its quantization bins. 
Control features are then generated by
argmax sampling over the estimated quantization bin probabilities. 
For training, we define the loss on weighted negative log probabilities: $\mathcal{L}_{\text{cl}} = \mathcal{L}_{F_0} + \mathcal{L}_{L} + \mathcal{L}_{P} + \mathcal{L}_{C}$, where 
\begin{align}
   \mathcal{L}_{F_0} &= -\sum_{i=1}^{n_{\text{strings}}}\sum_{t=1}^{T} \sum_{b=1}^{n_{\text{$f_0$ bins}}} {F_0}(i,t,b) \log(\hat{F_0}(i,t,b)) \cdot \bm{l}(i,t) \cdot \bm{p}(i,t) \\
   \mathcal{L}_{L} &= -\sum_{i=1}^{n_{\text{strings}}}\sum_{t=1}^{T} \sum_{b=1}^{K} {L}(i,t,b) \log(\hat{L}(i,t,b)) \\
   \mathcal{L}_{P} &= -\sum_{i=1}^{n_{\text{strings}}}\sum_{t=1}^{T} \sum_{b=1}^{K} {P}(i,t,b) \log(\hat{P}(i,t,b)) \cdot \bm{l}(i,t) \\
   \mathcal{L}_{C} &= -\sum_{i=1}^{n_{\text{strings}}}\sum_{t=1}^{T} \sum_{b=1}^{K} {C}(i,t,b) \log(\hat{C}(i,t,b)) \cdot \bm{l}(i,t). 
\end{align}
\texttt{ctr-syn-cl} has 72.3M parameters with 54M coming from $\mathcal{C}_{\text{cl}}$.

\vspace{-2mm}
\subsection{Joint training of control and synthesis sub-modules}
Past work has shown that joint training of sub-modules can lead to better performance for neural waveform synthesis of piano sounds \cite{shi_can_2023}. We therefore propose a system where the control model is jointly trained together with a pre-trained synthesis model.
We refer to this system as \texttt{ctr-syn-jt}.
The new control model is written: 
$\hat{F_0},\bm{\hat{l}},\bm{\hat{p}},\bm{\hat{c}} = \mathcal{C}_\text{jt} (X_{\text{pitch}}, X_{\text{vel}}, s)$.

During joint training, we generate a predicted F0 contour $\hat{\bm{f_0}}$ by argmax sampling of $\hat{F_0}$, which is passed to the synthesis model to generate a waveform $\hat{y}$.
The loss used during the joint training is $\mathcal{L}_{\text{jt}} = \mathcal{L}_{F_0} + \text{MSSL}(y,\hat{y})$,
where $y$ is the natural microphone audio and
$\hat{y}$ is the audio waveform synthesized from predicted control features.
The loss $\mathcal{L}_{\text{jt}}$ does not apply direct supervision to the predictions of $\bm{p}$, $\bm{l}$ and $\bm{c}$ to give the model flexibility in terms of the information passed from the control model. 
However, we keep supervision of F0 since it has been shown to be difficult to tune the frequency of an oscillator with gradient descent using a point-wise spectral loss.\cite{turian_im_2020,hayes_sinusoidal_2023}.
\texttt{ctr-syn-jt} has 72.2M parameters in total with 53.9M coming from $\mathcal{C}_{\text{jt}}$.

\vspace{-2mm}
\subsection{Unified model}

\begin{figure}
    \centering\includegraphics[width=\columnwidth]{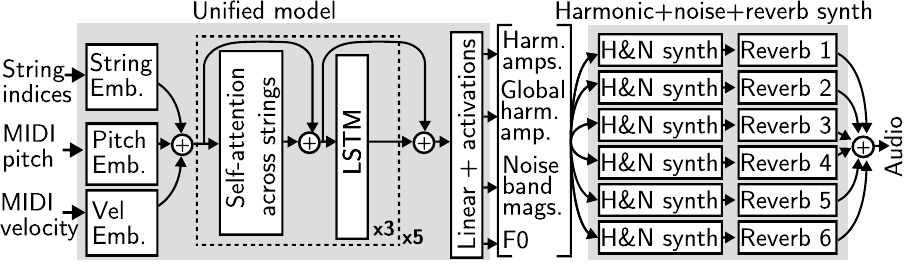}
    \caption{Details of the unified model architecture.}
    \label{fig:unified}
    \vspace{-5mm}
\end{figure}

Following promising results from initial experiments with joint training of control and synthesis model, we simplify our approach by merging the control model and synthesis decoder into a single network $\mathcal{U}$ that predicts synthesis parameters directly from the MIDI input and is shown in Figure \ref{fig:unified}. 
We refer to this system as \texttt{unified}. We write $\mathcal{U}$ as
$\hat{F_0}, H, \bm{a}, N = \mathcal{U}(X_{\text{pitch}}, X_{\text{vel}}, s)$.

The unified model is trained jointly with the trainable reverb to minimize the sum of the multi-scale spectral loss and the F0 classification loss $\mathcal{L}_u=\mathcal{L}_{F_0} + \text{MSSL}(y,\hat{y})$.
\texttt{unified} has 89.8M parameters with 89.7M coming from $\mathcal{U}$.

\section{Experiments}
\subsection{Experimental conditions}

\noindent
\textbf{Dataset:}
We use the \emph{GuitarSet} dataset \cite{xi_guitarset_2018},
which contains audio and pitch annotations for 360 acoustic guitar performances, totalling just over 3 hours. 
GuitarSet encompasses six different guitar players each playing solo and accompaniment for three chord progressions in five different styles in two different tempos in random keys.
The performances are recorded with both a microphone as well as a hexaphonic pickup.
A hexaphonic pickup is a guitar pickup that produces one audio channel for each guitar string.
It is important to note, however, that the correspondence between channels and strings is imperfect as the pickups also capture some signal from neighbouring strings. 
This phenomenon is called bleed.
To reduce bleed, the GuitarSet authors provide hexaphonic recordings processed with the KAMIR \cite{pratzlich_kernel_2015} bleed removal algorithm. 
MIDI pitch annotation, extracted semi-automatically from the hexaphonic recordings, is included in GuitarSet. Importantly, the pitches of the MIDI pitch annotation are not quantized to semitones.
Since velocity annotation is not included in GuitarSet, we use the peak unit scaled dB(A) loudness of every note as a proxy for MIDI velocity \cite{jonason_control-synthesis_2020}.

We split the data such that
no recordings from the same performer-progression-style triple occur in both partitions.
Splitting it randomly instead risks having recordings of a specific performer playing the same progression in the same style appearing in both test and development splits.
To limit author influence on which recordings end up in each split, we assign random aliases to each player, progression and style prior to splitting.
Using the aliases, we select 36 recordings for testing while balancing for diversity of players, progressions and styles.
We use the remaining 324 recordings for model development, from which we randomly choose 306 for training, and keep the remaining 18 recordings for validation.

\noindent
\textbf{Feature extraction:} Ground truth values for the control features are extracted from KAMIR-processed hexaphonic recordings \cite{xi_guitarset_2018}.
We compute F0 and periodicity using a PyTorch implementation of \emph{CREPE}  \cite{kim_crepe_2018,morrison_torchcrepe_2022} and scale F0 with MIDI spacing \cite{engel_ddsp_2019} to $[0,1)$ where $0$ corresponds to ($35Hz$) and $1$ is ($1200Hz$).
We measure loudness with A-weighting using \texttt{librosa} \cite{mcfee_librosa_2015} and scale it to $[0,1)$ where $0$ is $-80$ dB and $1$ is $0$ dB \cite{engel_ddsp_2019}. 
Finally, we compute spectral centroid with librosa \cite{mcfee_librosa_2015} and divide it by the Nyquist frequency.

\noindent
\textbf{Model training and inference:} 
All training uses the ADAM optimizer with $\beta_1=0.99$ and $\beta_2=0.999$, a learning rate decay of $0.99$ and early stopping with a patience of $5$ epochs.
We train one synthesis model for \texttt{ctr-syn-rg}, \texttt{ctr-syn-cl} and \texttt{ctr-syn-jt}, using a learning rate of $3\times10^{-4}$.
Control models for \texttt{ctr-syn-rg} and \texttt{ctr-syn-cl} are trained using a learning rate of $1\times10^{-4}$. 
Joint training of control and synthesis model in \texttt{ctr-syn-jt} uses a learning rate of $1\times10^{-4}$.
The \texttt{unified} system's training uses a learning rate of $1\times10^{-4}$. 
Training excerpts are 8 seconds in duration, extracted from recordings at random time offsets.
Both input features and control features have a feature frame rate of $128$ Hz.
Audio is generated at $48$ kHz. Computing the MSSL uses window sizes $[192, 384, 768, 1526, 3072, 6144, 12288]$.
 In order to render full recordings from the test set, we window the conditioning into 8-second windows with a 4-second skip length and mix the resulting windowed audio with 100 ms linear crossfade starting 2 seconds into the preceding window. 

\vspace{-2mm}
\subsection{Evaluation}

We now perform objective and subjective evaluation of the proposed systems, and also include audio produced from a pre-trained synthesis model given ground truth control features, denoted as \texttt{oracle-syn}.
The subjective evaluation also includes natural audio as well as renders of the MIDI with the free commercial sample-based guitar synthesizer \texttt{Ample Guitar lite} \cite{amplesound_ample_2015}.

\vspace{1mm}
\noindent
\textbf{Objective evaluation:}
While rendering the test samples, we compute the MSSL between natural and synthesized audio for each 8-second window. The average MSSL across all 8-second clips for all recordings is shown in Table \ref{tab:evaluation}.
Out of the proposed systems, we find that \texttt{ctr-syn-rg} shows the worst spectral loss 
whereas the best spectral losses are seen for \texttt{ctr-syn-jt} and \texttt{unified}.
However, the MSSL for \texttt{oracle-syn} is much lower than that of our best proposed systems.

\begin{figure}
    \centering
    \includegraphics[width=\columnwidth]{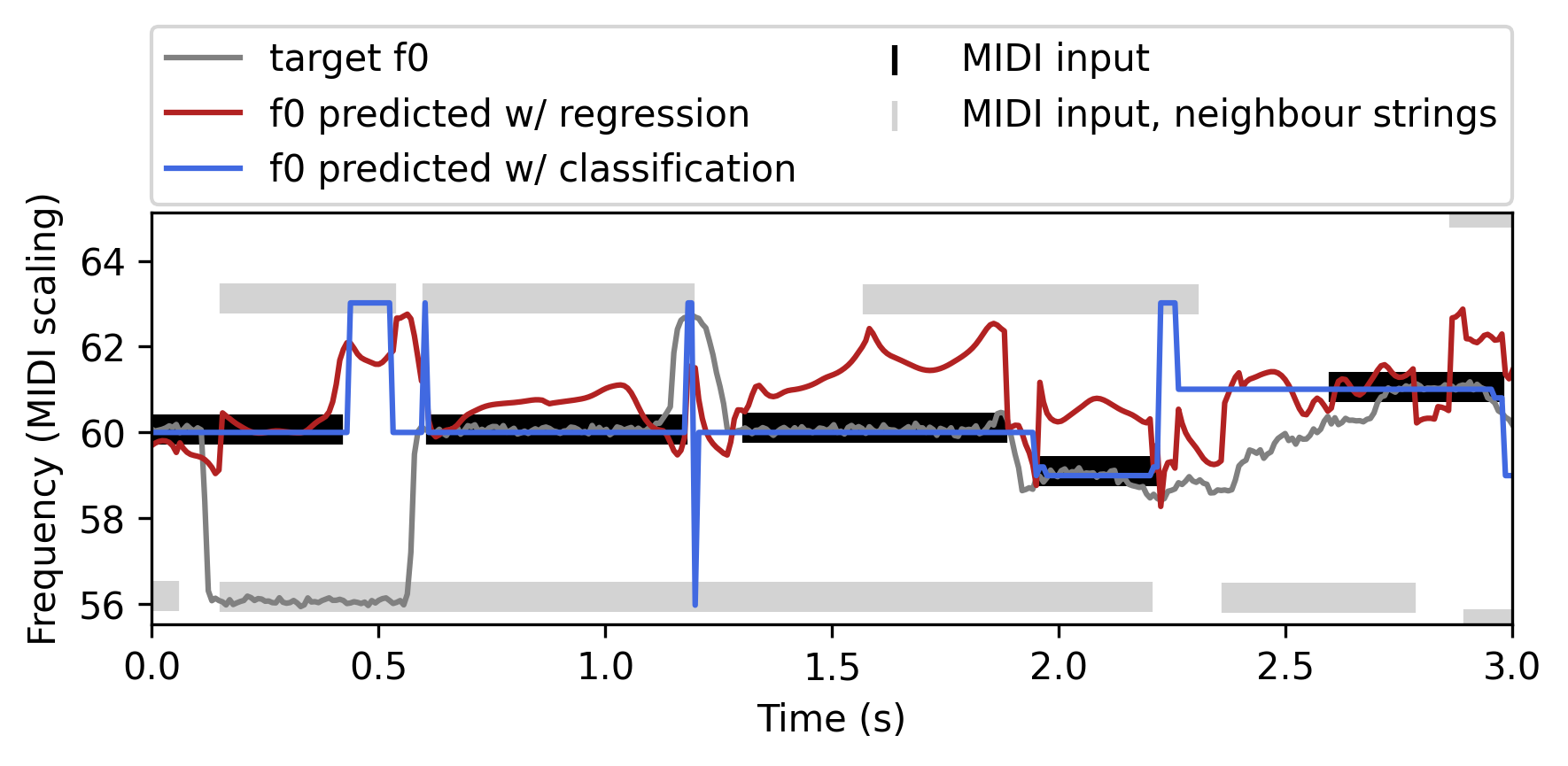}
    \caption{Fourth string predicted F0 from \texttt{ctr-syn-rg} and \texttt{ctr-syn-cl} against the target F0 (CREPE) and input MIDI pitch. Also shown is MIDI input from strings 3 and 5. In the first half-second we see the target F0 erroneously jump to the pitch of the 3rd string. We also see high inaccuracy in the regression-based F0 prediction with respect to both target F0 and MIDI input pitch.
}
    \label{fig:f0-prediction}
\end{figure}

To evaluate pitch accuracy, we follow a series of steps. We first extract F0 estimates for each string from the synthesized string channels using CREPE,
and then quantize them into semitones. Subsequently, we filter out frames in which there are no active input MIDI notes on the corresponding string. Finally, we compare these semitone estimates to two reference sources: the string-wise semitone-quantized F0 values estimated by CREPE from the natural hexaphonic audio and the string-wise semitone-quantized input MIDI pitch data.
The resulting average accuracies, denoted \texttt{CREPE acc.} and \texttt{MIDI acc.} are shown in Table \ref{tab:evaluation}. We notice that \texttt{ctr-syn-rg}, which predicts F0 with regression, performs poorly in pitch accuracy while systems that predict F0 with classification (\texttt{ctr-syn-rg}, \texttt{ctr-syn-cl},\texttt{unified}) obtains far higher pitch accuracies.
Based on visual inspection of F0 predictions, shown in Figure \ref{fig:f0-prediction}, we hypothesize that the failure of regression to accurately predict F0 originates from errors in the target F0 contours caused by bleed from neighbouring strings. In attempting to minimize the MSE to the flawed target, the regression model places its prediction between the pitch of the target string and the active pitches from neighbouring strings.
We also observe that the classification-based F0 predictions are closer in semitone accuracy to the MIDI pitch input than they are to the CREPE contour, which they are trained to predict. One explanation for this could be that they learn to rely on naively forwarding the input MIDI pitch to the F0 predictions.

\noindent
\textbf{Subjective evaluation:} We conducted an online listening test. Since the durations of the full recordings (18 to 44 seconds) are considerably longer than what is typically presented in a listening test, we segment all audio into halves and quarters, giving us 216 segments with durations ranging from 4 to 22 seconds.

We recruited 66 unique listeners who self-reported having experience playing guitar. Each listener rated either one or two sets of 108 samples, totalling 70 sets, balanced to have approximately the same number of samples for each system. The listeners were asked to rate the samples on a 5-point scale (very bad, bad, acceptable, good, very good), taking into account sound quality, naturalness and appropriateness of pitch. 
We perform a statistical analysis of the system scores using a two-sided Mann-Whitney U test with Bonferroni correction with $\alpha=0.05$.

\begin{table}[t]
\caption{Objective and subjective evaluation results.}
\label{tab:evaluation}
\centering
\small
\footnotesize
\begin{tabular}{lrrrrrr}
\toprule
system  & MSSL $\downarrow$ & CREPE acc.$\uparrow$ & MIDI acc.$\uparrow$  
& MOS$\uparrow$ \\
\midrule

\texttt{natural}  & - & - & - 
& 4.10 \\
\texttt{ample-gtr} & - & - & - &  4.08 \\

\texttt{oracle-syn} & 6.21 & 0.94&	0.94 &   3.08 \\

\midrule

\texttt{ctr-syn-rg} & 10.51 & 0.33	&0.34 &  1.18 \\
  \texttt{ctr-syn-cl} & 8.02 & 0.89 &	0.96 &  2.64 \\
\texttt{ctr-syn-jt} &      \textbf{7.64}  &0.89&	0.97 &  3.00 \\
   \texttt{unified} &  7.71  & \textbf{0.90} &	\textbf{0.97} &    \textbf{3.38}  \\

\bottomrule
\end{tabular}
\vspace{-5mm}
\end{table}

The last column of Table \ref{tab:evaluation} shows the mean opinion scores (MOS) of the systems.
According to the two-sided Mann-Whitney U test with Bonferroni correction, there are statistically significant differences in the MOS ratings among all pairs of systems, except for two pairs: \texttt{natural} and \texttt{ample-gtr}, as well as \texttt{oracle-syn} and \texttt{ctr-syn-jt}.
We find that \texttt{ctr-rg-syn} performs the worst of all systems with an overall MOS of $1.18$. The next best is \texttt{ctr-rg-syn} with an overall MOS of $2.64$. After that \texttt{ctr-syn-jt} obtains a MOS of $3.00$. Finally, \texttt{unified} performs the best out of all four proposed systems with a MOS of $3.38$, even outperforming \texttt{oracle-syn}.
This is lower than both \texttt{ample-gtr} and \texttt{natural}, which obtain a MOS of $4.08$ and $4.10$. 

These indicate, first, that all of the proposed improvements have resulted in audible improvements of the synthesized guitar performance. Second, it is also intriguing that \texttt{unified} had a higher MOS rating than \texttt{oracle-syn}. A deeper analysis of this observation is needed in the future. Finally, although the proposed \texttt{unified} system was not as good as the sample-based system \texttt{ample-gtr}, we can see that it was able to synthesize guitar sounds of reasonable quality despite being trained on only a few hours of data.

\section{Conclusion}
We have developed four different DDSP-based neural waveform synthesis systems for polyphonic guitar performance from string-wise MIDI input, and compared them with both objective metrics and subjective evaluation against natural audio and a sample-based baseline.
We find that switching from control feature regression to control feature classification improves performance considerably. Furthermore, we find that joint training of control and synthesis sub-modules further improves the synthesis quality. We have also suggested a simplification of the control synthesis architecture, which further improves synthesis quality.
Although the proposed systems were not as good as the sample-based baseline, our \texttt{unified} system is able to synthesize guitar performance of reasonable quality with only a few hours of data. 
Our next important steps will be taking full advantage of the benefits of neural methods, such as self-supervised pre-training with large amounts of unlabeled data.

\section{Acknowledgments}
This work is supported by MUSAiC: Music at the Frontiers of Artificial Creativity and Criticism (ERC-2019-COG No. 864189) and JST CREST Grants (JPMJCR18A6 and JPMJCR20D3) and MEXT KAKENHI Grants (21K17775, 21H04906, 21K11951, 22K21319).

\label{sec:pagestyle}
\bibliographystyle{IEEEbib}
\bibliography{references2}

\end{document}